\documentclass[aps,twocolumn,english,superscriptaddress,citeautoscript,preprintnumbers,amsmath,amssymb,floatfix,footinbib]{revtex4-1}
\usepackage[breaklinks=true,colorlinks,citecolor=blue,linkcolor=blue,urlcolor=blue]{hyperref}
\usepackage{amsmath}
\usepackage{graphicx}
\usepackage{dcolumn}
\usepackage{float}
\usepackage[utf8]{inputenc}
\usepackage{color}
\usepackage{soul}
\begin{document}
	
\title{Large nonsaturating magnetoresistance, weak anti-localization and non-trivial topological states  in SrAl$_2$Si$_2$}
\author{Sudip Malick}
\affiliation{Department of Physics, Indian Institute of Technology, Kanpur 208016, India}
\author{A. B. Sarkar}
\affiliation{Department of Physics, Indian Institute of Technology, Kanpur 208016, India}
\author{Antu Laha}
\affiliation{Department of Physics, Indian Institute of Technology, Kanpur 208016, India}
\author{M. Anas}
\affiliation{Department of Physics, Indian Institute of Technology Roorkee, Roorkee 247667, India}
\author{V. K. Malik}
\affiliation{Department of Physics, Indian Institute of Technology Roorkee, Roorkee 247667, India}
\author{Amit Agarwal}
\email{amitag@iitk.ac.in}
\affiliation{Department of Physics, Indian Institute of Technology, Kanpur 208016, India}
\author{Z. Hossain}
\email{zakir@iitk.ac.in}
\affiliation{Department of Physics, Indian Institute of Technology, Kanpur 208016, India}
\author{J. Nayak}
\email{jnayak@iitk.ac.in}
\affiliation{Department of Physics, Indian Institute of Technology, Kanpur 208016, India}

\begin{abstract}
We explore the electronic and topological properties of single crystal SrAl$_2$Si$_2$ using magnetotransport experiments in conjunction with first-principle calculations. 
We find that the temperature-dependent resistivity shows a pronounced peak near 50 K. We observe several remarkable features at  low temperatures such as large non-saturating magnetoresistance, Shubnikov-de Haas oscillations and cusp-like magneto-conductivity. The maximum value of magnetoresistance turns out to be 459\% at 2 K and 12 T. The analysis of the cusp-like feature in magneto-conductivity indicates a clear signature of weak anti-localization. Our Hall resistivity measurements confirm the presence of two types of charge carriers in SrAl$_2$Si$_2$, with low carrier density.
\end{abstract}
	\maketitle
	
\section{INTRODUCTION}
The new paradigm of topology in crystalline materials has brought relativistic physics to tabletop experiments. Current understanding of topological band theory and crystalline symmetries has led to the realization of several topological materials with novel bulk and surface properties, giving rise to new physics and functionalities. In particular, their exotic bulk and surface electronic structures can give rise to interesting transport phenomena such as extremely large magnetoresistance (MR), chiral anomaly, quantum oscillation, weak anti-localization (WAL) effect and ultra high mobility \cite{RevModPhys.82.3045, RevModPhys.90.015001,Liu864, MoscaConte2017, Nayak2017, Bian2016, ZrSiS, Kuroda2017,Li2019, YbCdSn}. Topological semimetals (TSM) like Dirac \cite{PhysRevB.100.195134,PhysRevLett.121.226401}, Weyl and nodal line semimetals \cite{PhysRevB.98.085122} are the consequence of such non-trivial states in which the bulk bands host symmetry protected gapless states between the valence and conduction bands near the Fermi level.

Recently, materials (122 phase) belonging to the space group $P\bar{3}m1$ (No. 164) have attracted significant interest owing to the existence of diverse topological phases. EuCd$_2$As$_2$ and EuCd$_2$Sb$_2$ are the two specific systems in this group, where magnetism plays a crucial role in defining their electronic and topological ground states. Depending on the magnetic spin configuration, these systems can either be topological insulators or Dirac semimetals \cite{PhysRevB.99.245147, Maeaaw4718, PhysRevB.100.201102, doi:10.1063/1.5129467, PhysRevB.97.214422}. On the other hand, a non-magnetic compound in the same group - CaAl$_2$Si$_2$, is predicted to host multiple topological states such as topological Dirac nodal lines and type-II Dirac fermions. The inclusion of spin-orbit coupling (SOC) in this system, opens up a small gap in the Dirac line nodes, while the crystal symmetries protect  the Lorentz symmetry violating tilted Dirac point states \cite{PhysRevB.101.205138}. Recently, the topological Dirac semimetallic phase in CaAl$_2$Si$_2$ has been identified experimentally by Deng {\it et. al.} \cite{PhysRevB.102.045106}.

Here, we have explored another non-magnetic system from the same space group ($P\bar{3}m1$) - SrAl$_2$Si$_2$. We report the synthesis of single crystal of SrAl$_2$Si$_2$ and present a systematic study of its physical properties by using magnetotransport measurements in conjunction with first-principle calculations. Our magnetotransport measurements show i) a broad peak in the temperature-dependent resistivity, ii) large non-saturating MR upto $B=12$ Tesla, iii) quantum oscillations in the conductivity, iv) weak anti-localization, and v) two carrier dependent Hall resistivity which is non-linear in the magnetic field strength. Our density functional theory based calculations clearly indicate the presence of a Dirac point in the electronic dispersion, the existence of interesting surface states at the Fermi energy and multiple electron and hole pockets at the Fermi energy - consistent with the observed  Hall resistivity. 

\section{METHODS AND CRYSTAL STRUCTURE}
Single crystals of SrAl$_2$Si$_2$ were synthesized by the standard flux method with Al being used as a flux \cite{PhysRevB.101.205138}. High purity elements Sr (99.99\%, Alfa Aesar), Al (99.999\%, Alfa Aesar) and Si (99.99\%, Alfa Aesar) were taken in 1:20:2 molar ratio and mixed in an alumina crucible and then sealed in an evacuated quartz ampule. The whole assembly was heated to 1373 K, maintained at this temperature for 20 hours and then slowly cooled down to 1023 K at a rate of 1.5 K/hour. The excess Al flux was removed by centrifuging to extract the single crystals. The typical dimensions of the crystals were 1.5 mm$\times$1 mm$\times$0.5 mm. The crystal structure and the phase purity were determined by x-ray diffraction (XRD) in a PANalytical X'Pert PRO diffractometer with Cu K$\alpha_1$ radiation and by energy dispersion x-ray spectroscopy (EDS) in JEOL JSM-6010LA. Electrical resistivity and magnetotransport measurements were performed by the conventional four probe technique in a physical property measurement system (Dynacool PPMS, Quantum Design). 

\begin{figure}
	\includegraphics[width=8.5cm, keepaspectratio]{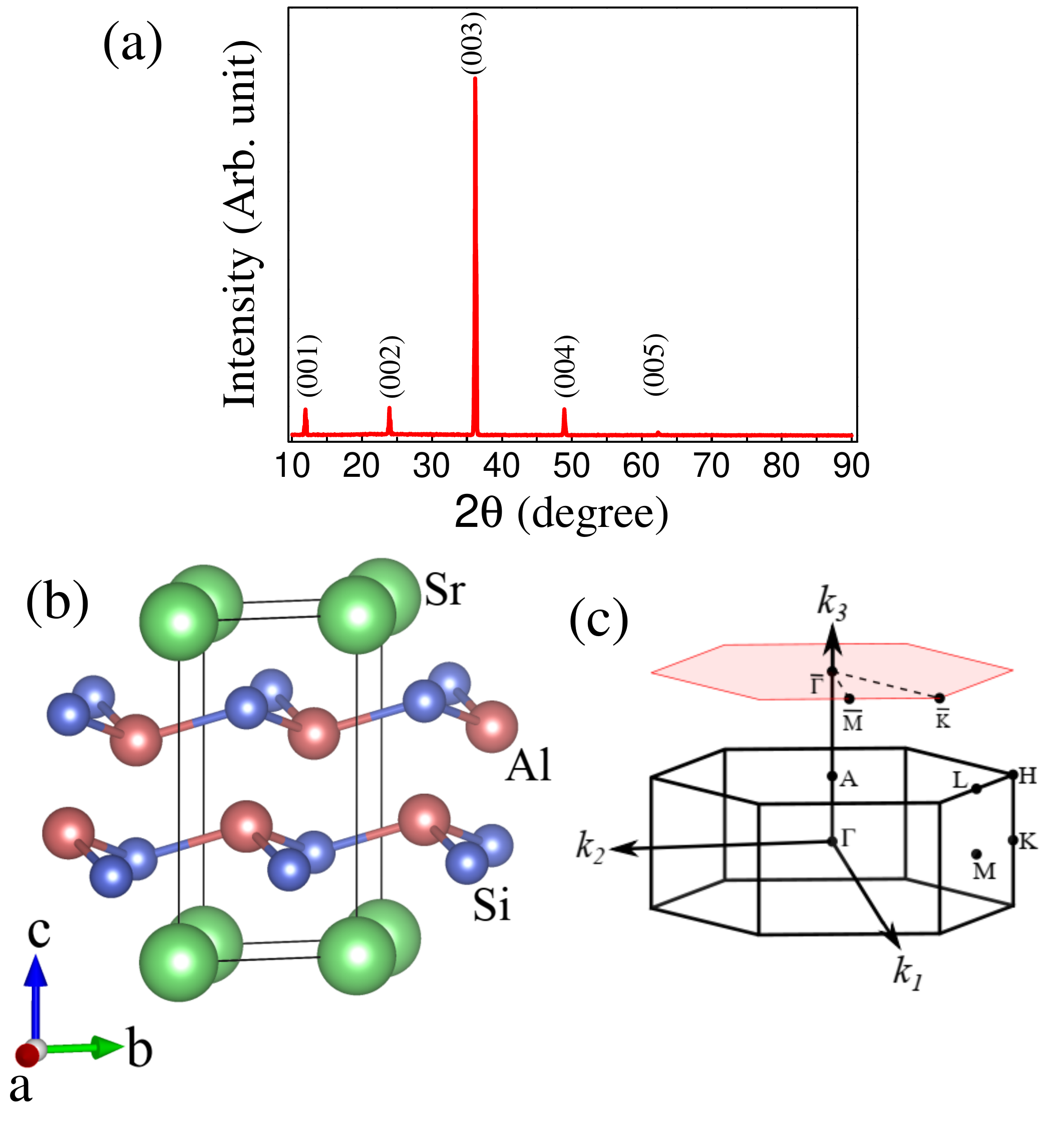}
	\caption{\label{Fig1} The crystal structure and Brillouin zone of SrAl$_2$Si$_2$. (a) Single crystal XRD pattern of SrAl$_2$Si$_2$ with sharp peaks confirm the good quality of the single crystals. (b) The schematic representation of the crystal structure, along with its unit cell. (c) Bulk and (001) projected hexagonal surface Brillouin zone, with the high symmetry points marked.}
\end{figure}

We calculated the electronic structure of SrAl$_2$Si$_2$ using the framework of density functional theory (DFT) as implemented in the Vienna {\it ab-initio} simulation package (VASP) \cite{kresse1996efficient, kresse1999ultrasoft}. The exchange-correlation effects were treated within the generalized gradient approximation (GGA) in the form of Perdew-Burke-Ernzerhof (PBE) type exchange correlation potential \cite{Perdew1996}. We used the kinetic energy cut-off for the plane-wave basis set as 500 eV and a $\Gamma$-centered $9\times9\times 9$  $k$-mesh  to perform the Brillouin zone (BZ) integration \cite{Monkhorst1976}. The tight binding model Hamiltonian was constructed using the VASP2WANNIER90 interface \cite{PhysRevB.56.12847}. We used the iterative Green's function method, as implemented in the WANNIERTOOLS package \cite{WU2017} to obtain the surface and bulk spectral function. We plotted the crystal structure using the VESTA software. 

The single crystal XRD pattern of SrAl$_2$Si$_2$ is presented in Fig.~\ref{Fig1}(a). It crystalizes in a hexagonal structure with space group $P\bar{3}m1$, as shown in Fig.~\ref{Fig1}(b). The obtained lattice parameters from the Rietveld refinement of the powder XRD data are $a = b = 4.190$~\AA ~and $c = 7.429$~\AA, which agree well with the previous report \cite{KAUZLARICH2009240}. SrAl$_2$Si$_2$ is a layered material with Si and Al atoms being arranged in a layer, followed by a layer of Sr atoms. The Sr atoms are situated at the Wyckoff site (0, 0, 0), whereas the Si and Al atoms are situated at (1/3, 2/3, 0.2762) and (1/3, 2/3, 0.6245), respectively. The structure possesses three and two-fold rotational symmetry along with three vertical mirror planes, $\mathcal{M}_{100}$, $\mathcal{M}_{010}$, and $\mathcal{M}_{110}$.  We show the bulk BZ along with the projected (001) surface BZ in Fig.~\ref{Fig1}(c).
    
\begin{figure*}[htb]
	\centering
	\begin{tabular}{@{}cccc@{}}
		\includegraphics[width=0.33\textwidth]{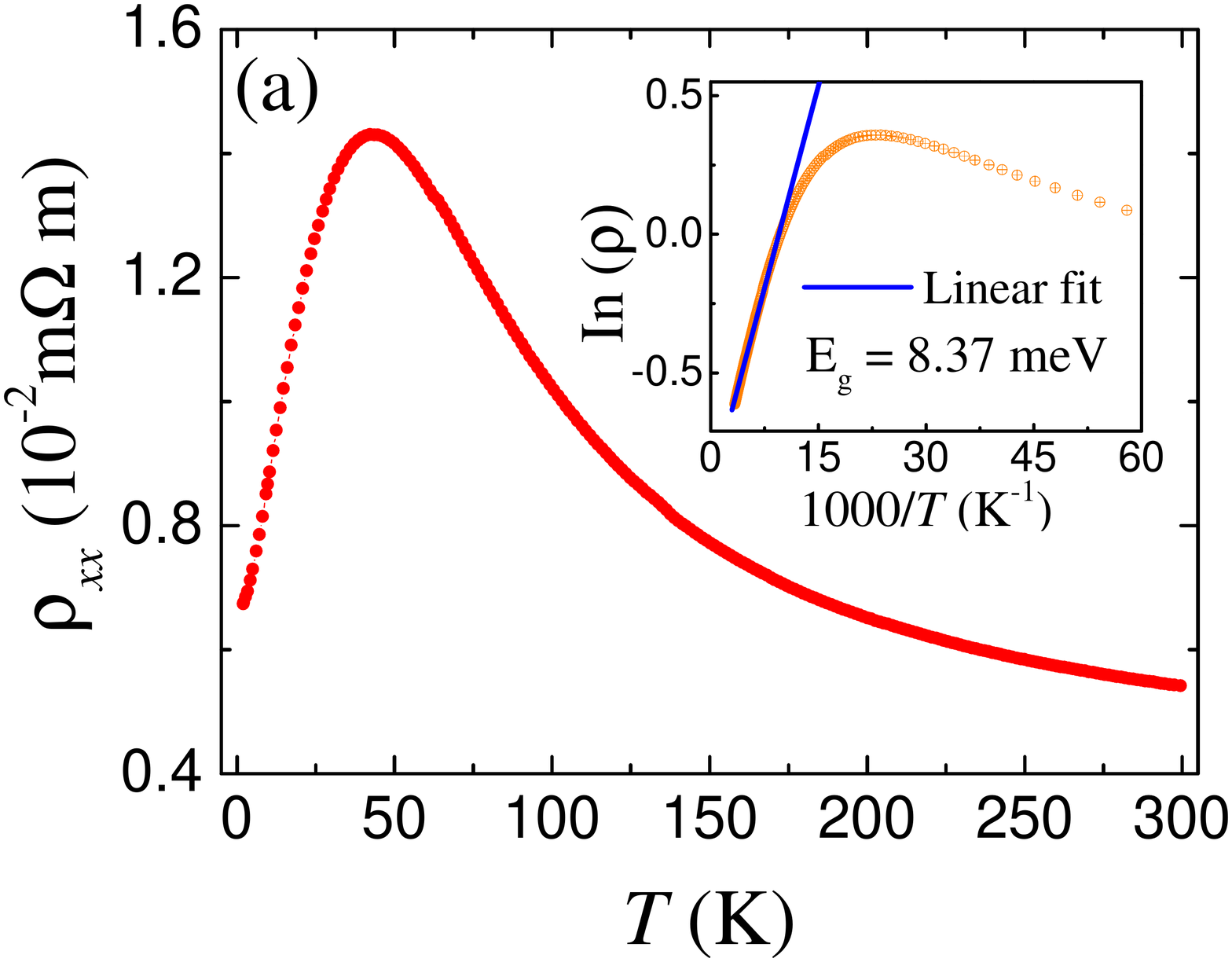} &
		\includegraphics[width=0.33\textwidth]{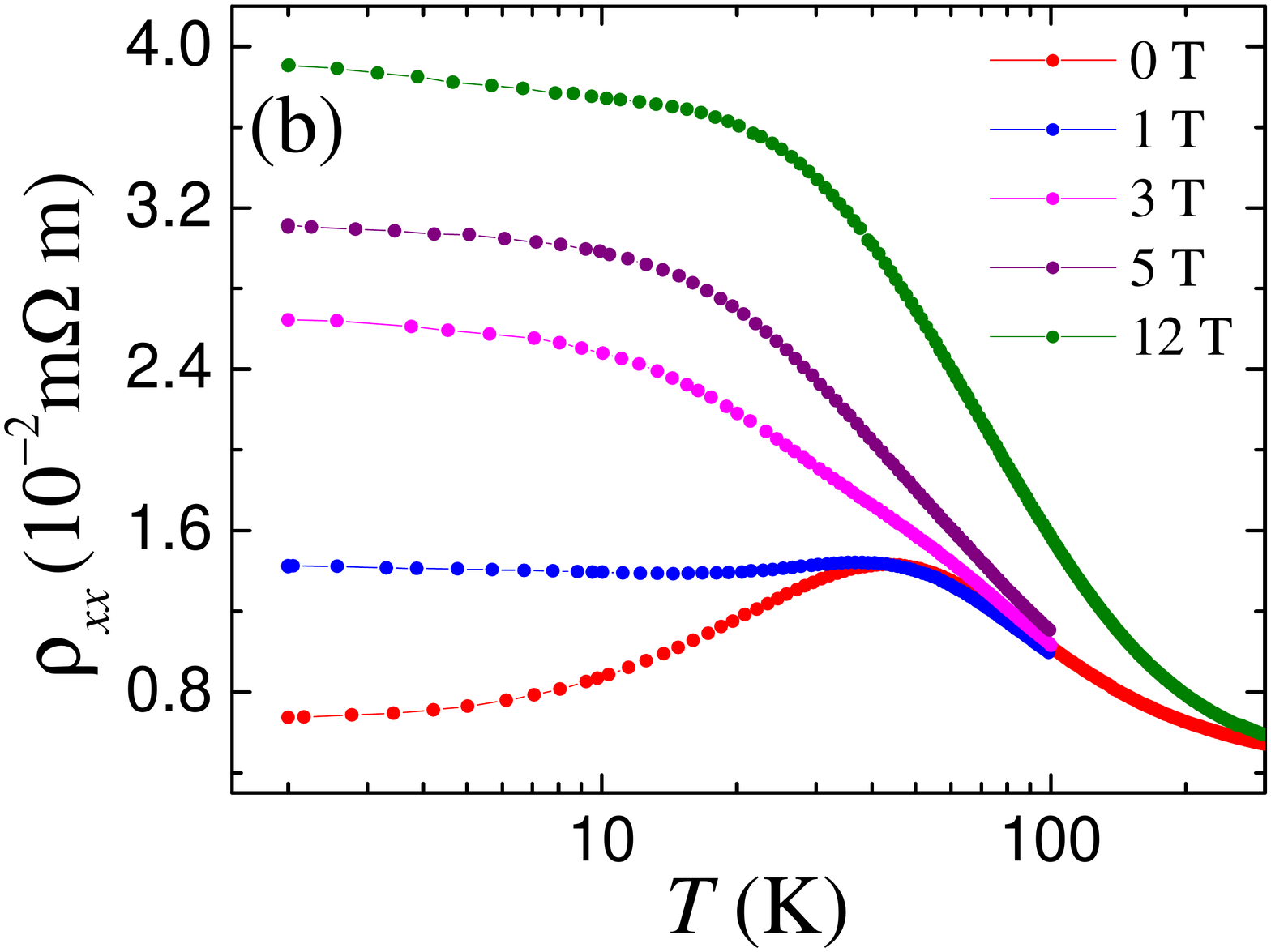} &
		\includegraphics[width=0.33\textwidth]{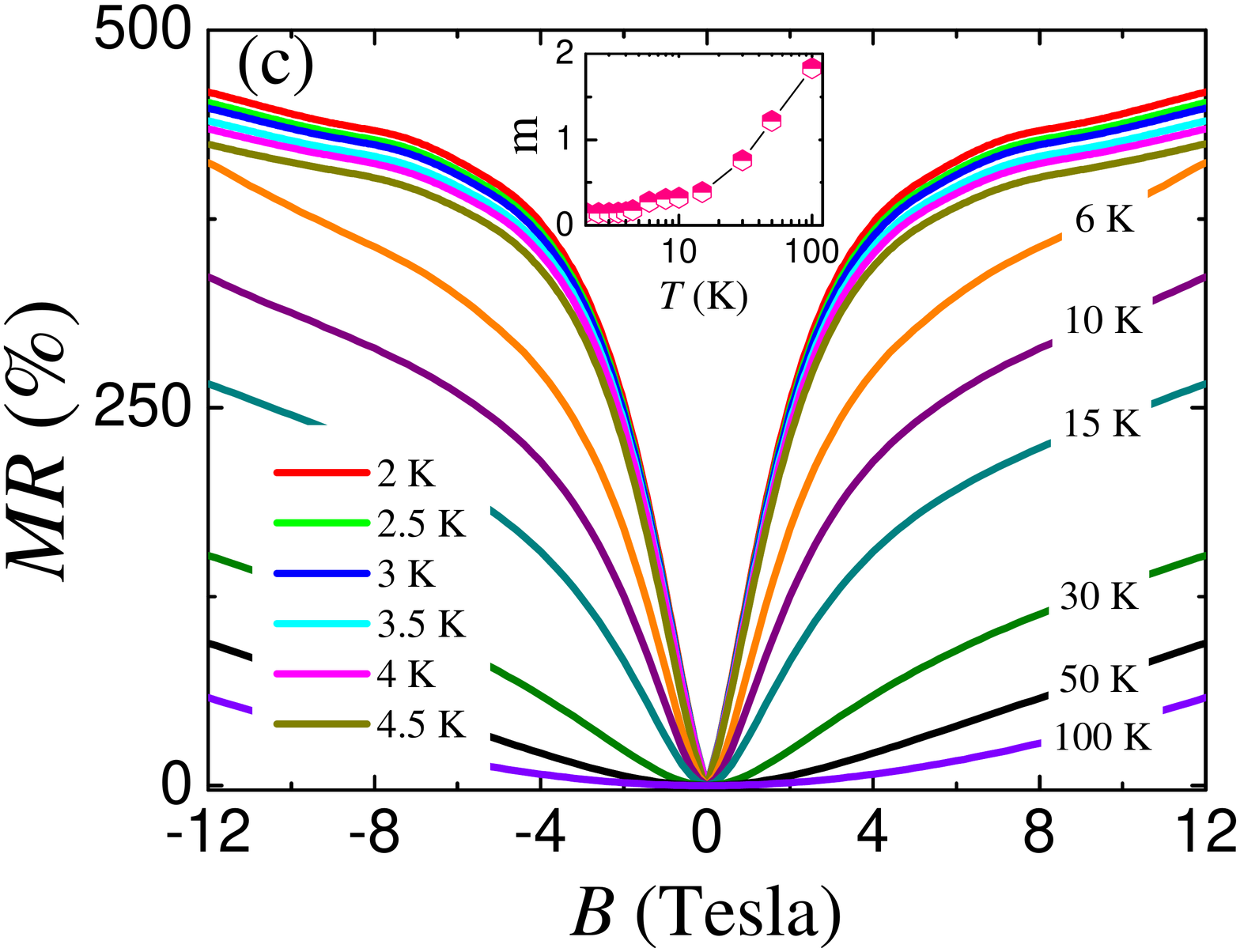} \\
		\includegraphics[width=0.33\textwidth]{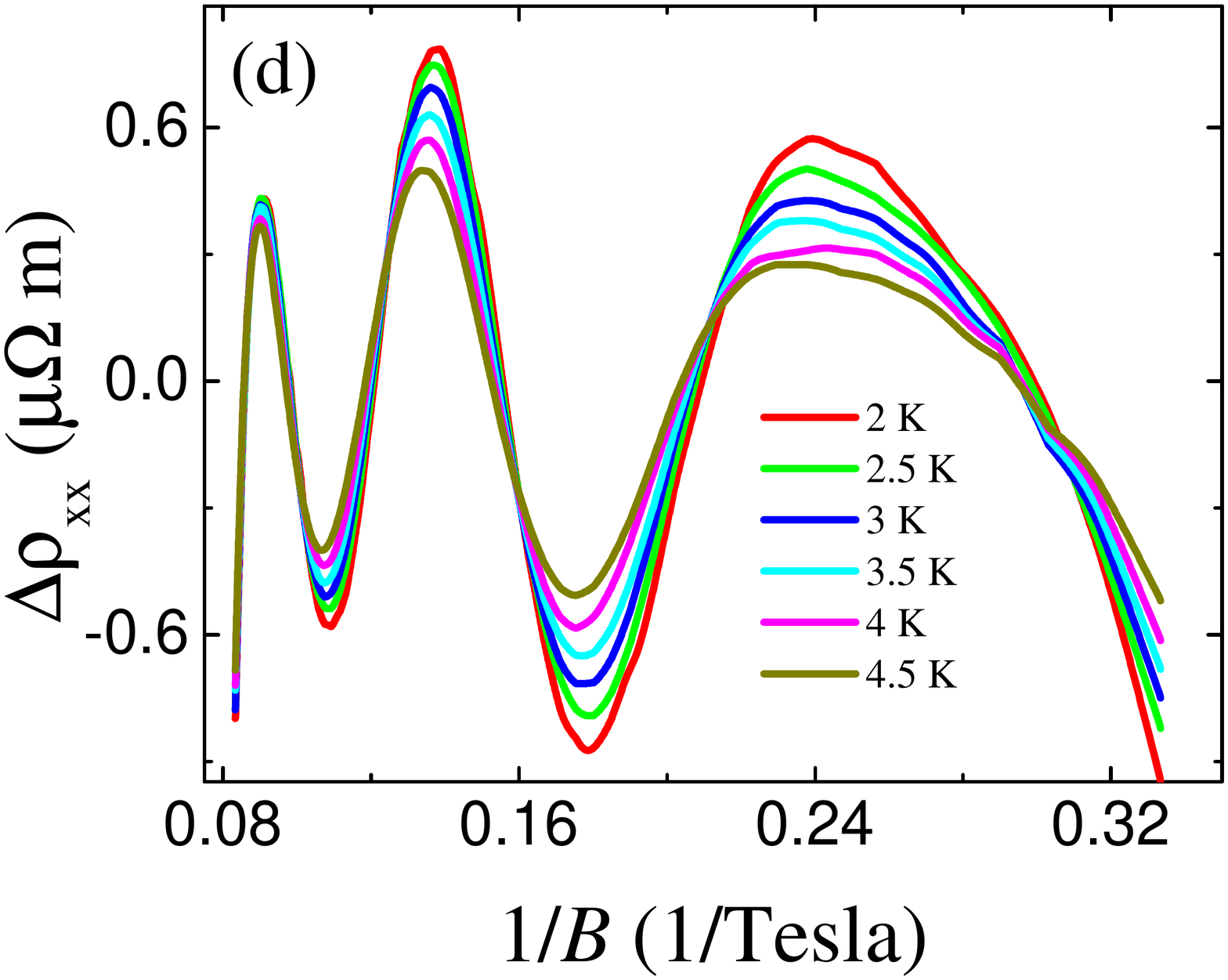} &
		\includegraphics[width=0.33\textwidth]{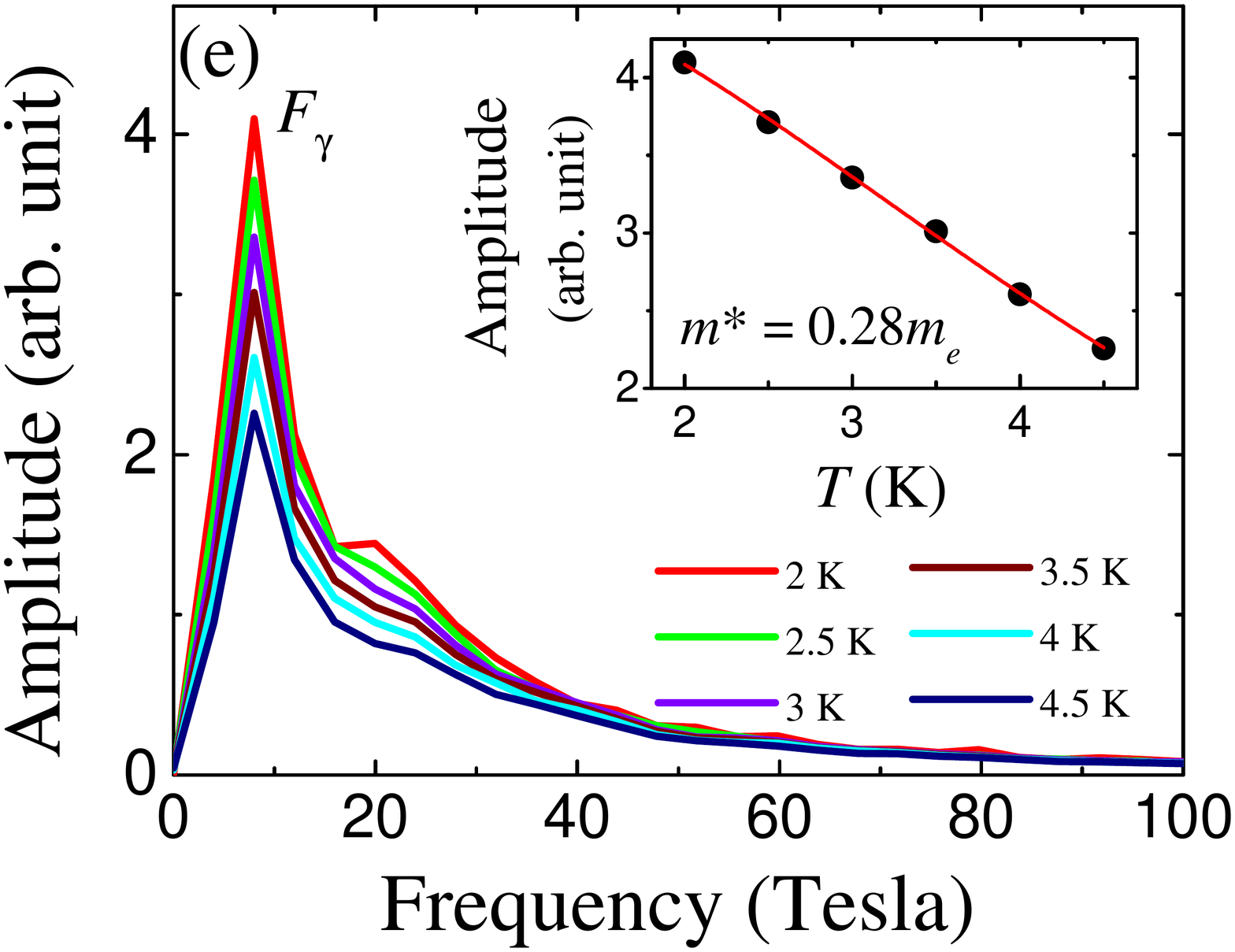}&
		\includegraphics[width=0.33\textwidth]{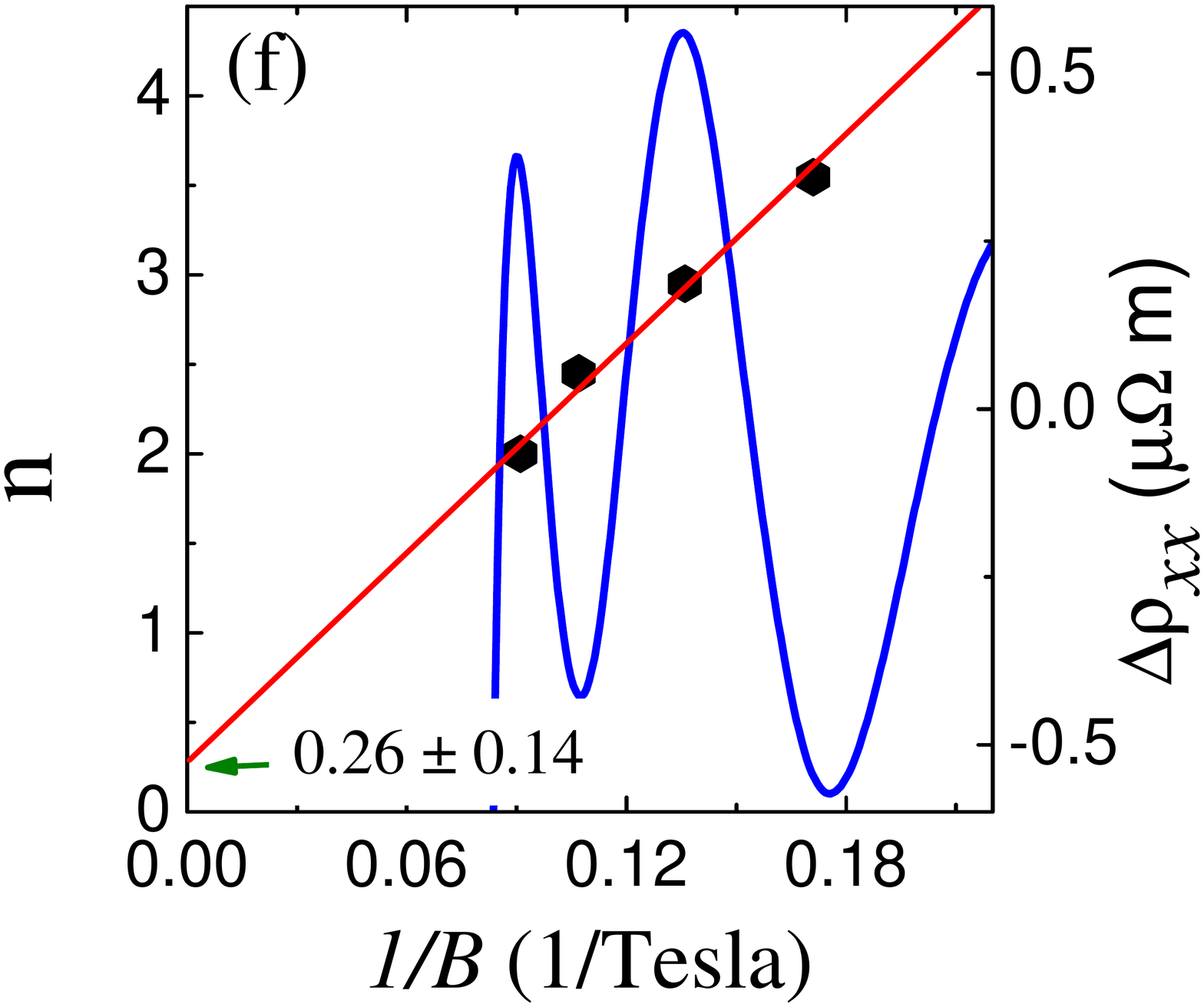}
			
	\end{tabular}
	\caption{\label{Fig2} Magnetotransport measurements of SrAl$_2$Si$_2$. (a) The temperature dependence of the longitudinal resistivity. The inset shows the fitting of the Arrhenius plot, with  an activation gap of $E_g = 8.4$ meV. (b) The temperature dependence of the longitudinal resistivity in the presence of a magnetic field. (c) MR as a function of $B \in [-12, 12]$ Tesla for different temperatures. The inset represents the dependence of the $m$ (MR=$\eta B^m$, where $\eta$ is a constant) with $T$. (d) The extracted oscillatory component of the SdH oscillations as a function of 1/$B$ for temperatures between 2 and 4.5 K. (e) The fast Fourier transform (FFT) spectra of the SdH oscillations indicate a fundamental frequency $F_{\gamma}$ = $8$ Tesla. The hump near 19 T may correspond to the second harmonic (f) The corresponding Landau level fan diagram for measurements at $T=4$ K.}
\end{figure*}

\section{Magnetotransport}

The measured temperature dependence of the longitudinal electrical resistivity (300 K to 2 K) is shown in Fig.~\ref{Fig2}(a). We find that the resistivity increases with the decrease of temperature from 300 K and it exhibits, a broad peak near 50 K, consistent with earlier reports \cite{KAUZLARICH2009240, PhysRevB.83.113105}. This is likely a consequence of the thermal activation of the carriers, which manifests as a semiconductor (or semimetal) like behavior at higher temperature \cite{LuPtSb}. To estimate the thermally activated effective energy gap or pseudogap, we have performed a linear fit to the resistivity in the Arrhenius plot above 100 K, as shown in the inset of Fig.~\ref{Fig2}(a), and estimated the `effective energy gap' to be $E_g = 8.4$ meV. Under the application of an external magnetic field, the low temperature metallic state is gradually suppressed  as shown in Fig.~\ref{Fig2}(b). Above a certain magnetic field ($B > 1$ T), only semiconducting-like behavior is observed with a resistivity plateau below 10 K. 

\subsection{Magnetoresistance}
Next, we present the MR as a function of the magnetic field at various temperatures in Fig.~\ref{Fig2}(c). The MR is defined by the following expression,
\begin{equation}
MR(\%)=\dfrac{\rho_{xx}(B)-\rho_{xx}(0)}{\rho_{xx}(0)}\times 100\%~.
\end{equation}
Here, $\rho_{xx}(B)$ and $\rho_{xx}(0)$ are the longitudinal resistivities in the presence and  absence of the external magnetic field, respectively. We find that SrAl$_2$Si$_2$ exhibits a large non-saturating MR. The MR increases to a value of 459\% at 2 K for 12 T. However, it decreases gradually with the increase of temperature and becomes 58\% at 100 K for 12 T. To understand the field dependency of the MR, we have fitted the MR curve at the high field region (6 T to 12 T) with the power law equation MR=$\eta B^m$, where $\eta$ is an arbitrary constant. The variation of $m$ with the temperature is shown in the inset of Fig.~\ref{Fig2}(c). At very low temperatures, we find $m$ to be nearly independent of temperature,  and then it increases  with temperature and reaches the value of 2 at \textit{T} = 100 K. A more thorough analysis of the MR data reveals two other distinct features at low temperatures: quantum oscillations at high magnetic field values and WAL at low magnetic field values. 

\subsection{Shubnikov-de Haas oscillations}
Quantum oscillations, namely the Shubnikov-de Haas (SdH) oscillations, are visible from the MR data at $B>3$ Tesla for $T\leq 4.5$ K.  To extract the oscillatory component ($\Delta\rho_{xx}$) of the SdH oscillation in $\rho_{xx}$, we have subtracted a smooth polynomial background from $\rho_{xx}(B)$. The plot of  $\Delta\rho_{xx}$  as a function of 1/$\it{B}$ is shown in Fig.~\ref{Fig2}(d). The Fourier transformation of the SdH oscillations, as shown in Fig.~\ref{Fig2}(e), reveals a fundamental frequency - $F_{\gamma}$= 8 T. The cross-sectional area ($A_F$) of the Fermi surface perpendicular to the applied field corresponding to this frequency is estimated using the Onsager relation $F= (\hbar/2\pi e)A_F$, where $\hbar$ is the reduced Planck's constant, and $e$ is the magnitude of electron charge \cite{ZrSiS}. The estimated area ($A_{F_\gamma}$) and carrier concentration ($n$) corresponding to the observed frequency  are 0.076 nm$^{-2}$ and 1.28$\times10^{17}$ cm$^{-3}$, respectively. To determine the effective  mass $m^{*}$ corresponding to the frequency $F_\gamma$, we have fitted the temperature dependent fast Fourier transform (FFT) amplitude [see inset of Fig. \ref{Fig2}(e)] with the thermal damping factor $R_T \propto \dfrac{2\pi^2k_BTm^*/\hbar eB}{sinh(2\pi^2k_BTm^*/\hbar eB)}$ of the Lifshitz-Kosovich formula, where $k_B$ is the Boltzmann constant \cite{shoenberg_1984}. The fitting yields $m^* = 0.28 m_e$, where $m_e$ is the free electron mass. A similar value of $m^{*}$ was found in the sister compound CaAl$_2$Si$_2$ \cite{PhysRevB.101.205138}.
 
For more insight into the nature of the bands forming the Landau levels (LL), we construct the Landau level fan diagram from the SdH oscillations in 4 K data by specifying the LL index as integer $n$ (half integer, $n+\frac{1}{2}$) against the conductivity maxima (minima) as shown in Fig.~\ref{Fig2}(f). We find an unequal spacing between the maxima and minima positions of the oscillation for different LLs in SrAl$_2$Si$_2$ [Fig.~\ref{Fig2}(d)]. A similar behavior is also found in other topological materials due to Zeeman splitting \cite{ZrSiS, Hu2016, Xing2020}, which leads to an unequal energy spacing between the different LL. One can estimate the nature of the bands that contribute to the SdH oscillations from the LL fan diagram using the Lifshitz-Onsager quantization condition, given by $A_F\frac{\hbar}{eB}=2\pi(n+\gamma)$ with $\gamma= \frac{1}{2}- \beta+\delta$ \cite{Busch2018}. Here, $2\pi\beta$ is the Berry phase which is zero for trivial band and $\pi$ for non-trivial band. The quantity $\delta$ is the additional phase which is zero for 2D system and $\pm\frac{1}{8}$ for 3D system. For a non-trivial system one would expect $\gamma= \pm 0.125$ and for trivial system $\gamma=$ 0.625 or 0.375 \cite{Busch2018, ROMANOVA201843, ZrSiS}. The intercept of linear fit to $n$ vs $1/B$ plot as shown in Fig. \ref{Fig2}(f) gives $\gamma= 0.26\pm0.14$, suggesting non-zero Berry phase ($0.45\pi \leq \phi_B \leq 1.01\pi$; considering standard error). The observed value of the Berry phase is like other non-trivial materials \cite{TiSb2,PtTe2, Cd3As2_SdH}.

\begin{figure}[t]
	\includegraphics[width=7.5cm, keepaspectratio]{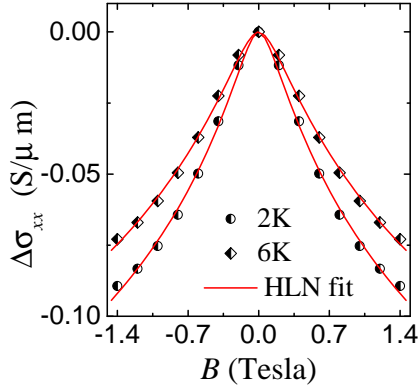}
	\caption{\label{Fig3} The magneto-conductivity as a function of $B$ for $T=2$ K and for $T=6$ K. The red solid line is the fit to the HLN  formula specified in Eq.~\eqref{eq2}.}
\end{figure}

\subsection{Weak anti-localization}
We now analyze another important feature observed in magnetotransport data - weak anti-localization (WAL). We find that WAL in SrAl$_2$Si$_2$ manifests at low magnetic field values for temperatures below $T=15$ K [see Fig. \ref{Fig2}(c)]. The WAL can be understood in terms of destructive quantum interference of electrons traversing different time reversal loops. The phase shifts of electron between two time reversed paths can lead to either constructive or destructive interference resulting in either weak localization or WAL, respectively. In a system with strong SOC or topologically protected bands, there is an additional $\pi$ phase shift in time reversed paths, which causes destructive interference and gives rise to negative magneto-conductivity (MC) \cite{BERGMANN1982815, BERGMANN19841, Hayasaka_2020}. The WAL correction in the MC is given by $\Delta \sigma_{xx}(B)= \sigma_{xx}(B)-\sigma_{xx}(0)$, where $\sigma_{xx}(B) = \rho_{xx}/(\rho_{xx}^2+\rho_{yx}^2) $ is the conductivity and $\rho_{yx}$ is the measured Hall resistivity. We have analyzed MC data using the well known Hikami-Larkin-Nagaoka (HLN) model. According to this model the MC correction arising from WAL can be described as \cite{LuPtSb, CaAuAs_SM}
 \begin{equation}
 \Delta \sigma_{xx}(B)= -A\left[ \Psi\left( \dfrac{1}{2}+\dfrac{\hbar}{4el^2_{\phi}B}\right) -ln\left(\dfrac{\hbar}{4el^2_{\phi}B}\right)\right].
 \label{eq2}
 \end{equation} 	
Here, $A=\frac{\alpha e^2}{\pi h}$, with the value of $\alpha = 1/2$ per conduction channel for 2D materials. In Eq.~\eqref{eq2}, $\Psi(x)$ is the digamma function and $l_{\phi}$ is the phase coherence length. We note that even though the HLN formula was derived for 2D materials, in practice it fits the data for 3D materials really well with large values of $\alpha$. We find that the MC data for 2 K and 6 K is well fitted with Eq.~(\ref{eq2}) in the low magnetic field (-1.4 T $\leq B \leq$ 1.4 T) regime, as shown in Fig.~\ref{Fig3}. The estimated value of $\alpha \sim 10^6$. Such a large $\alpha$-value was also found in several 3D materials due to the contribution of several conduction channels \cite{LuPtSb,ScPdBi,YbCdGe,CaAuAs_SM}. The obtained values of $l_{\phi}$ are 48.70 nm and 42.71 nm for 2 K and 6 K, respectively.

\begin{figure}
	\includegraphics[width=8.7cm, keepaspectratio]{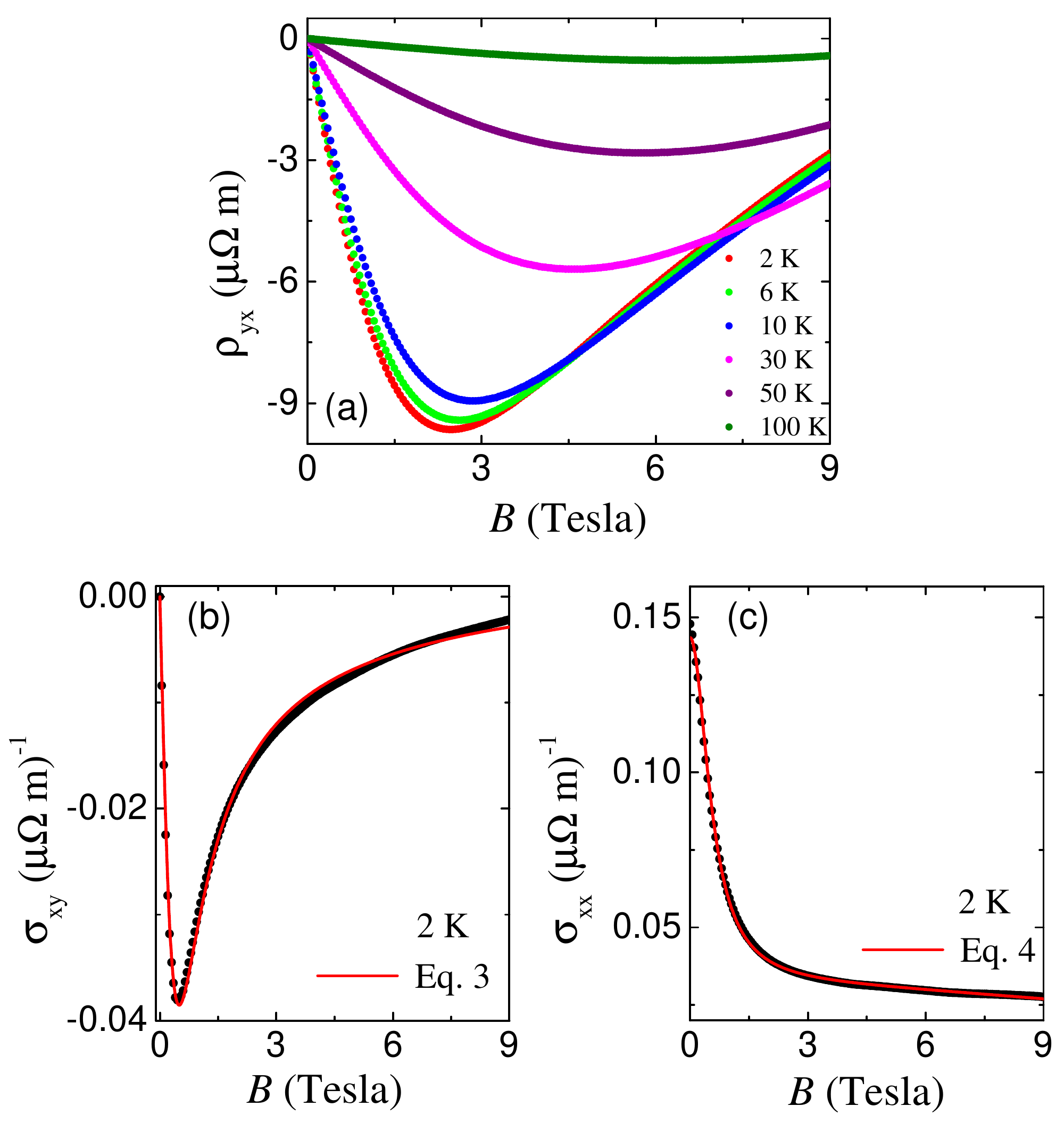}
	\caption{\label{Hall} (a) Hall resistivity measured at different temperatures. (b) Hall conductivity ($\sigma_{xy}$) and (c) longitudinal conductivity ($\sigma_{xx}$) as a function of the magnetic field at 2 K. The solid red line is the fitting of the two-band model as presented in Eq. \ref{sigma_xy} and Eq. \ref{sigma_xx}.}
\end{figure}

\subsection{Hall resistivity}
\label{Hall_resistivity}
\begin{figure}
	
	\includegraphics[width=8.6cm, keepaspectratio]{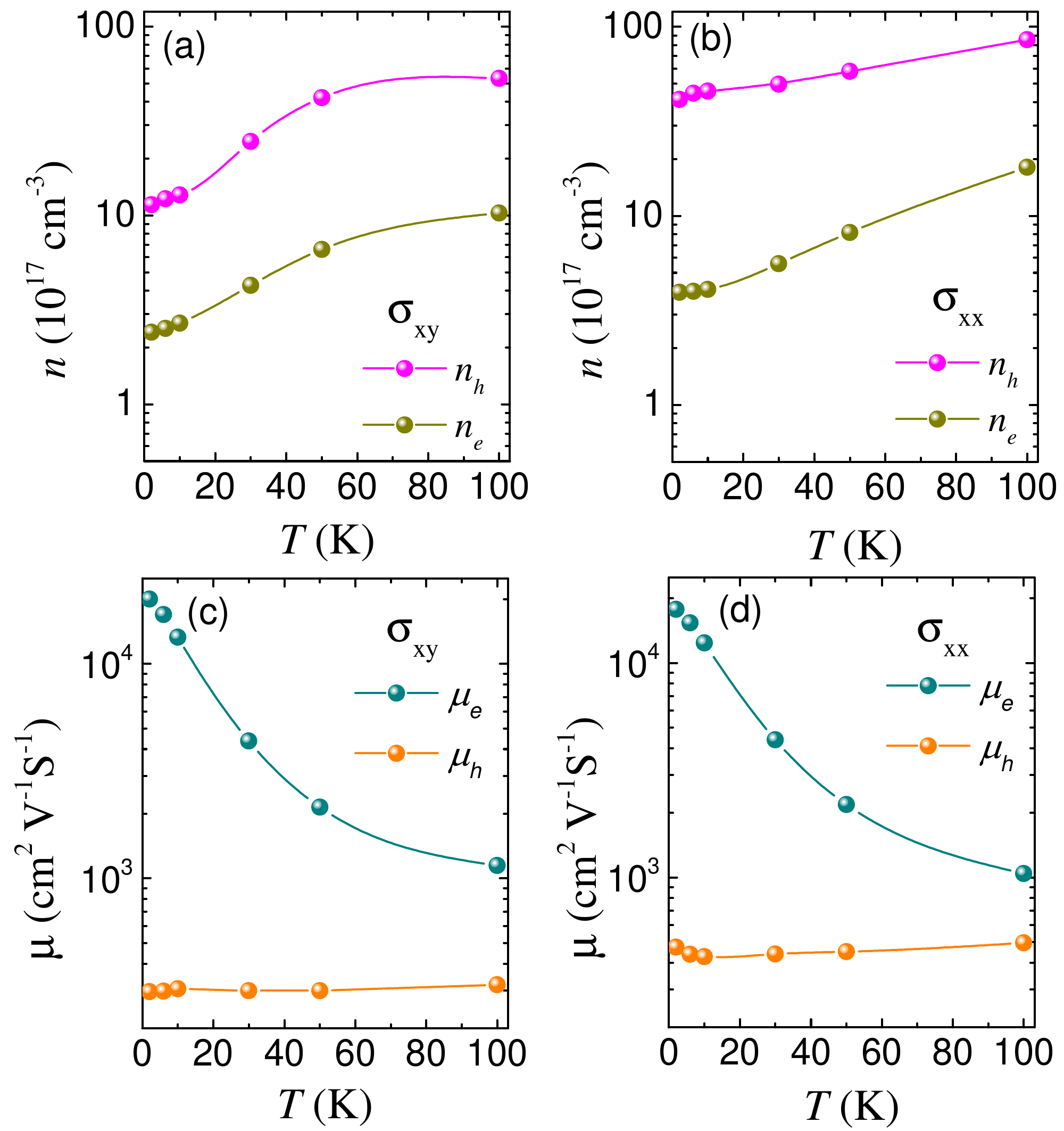}
	\caption{The variation of carrier concentration and mobility with temperature. (a) and (b) Estimated carrier concentrations from Eqs. \ref{sigma_xy} and \ref{sigma_xx}, respectively. (c) and (d) Calculated mobilities using Eqs. \ref{sigma_xy} and \ref{sigma_xx}, respectively.}
	\label{Hall_Num}
\end{figure}

To probe the details of the charge carriers participating in transport, we measured the Hall resistivity at different temperatures. The contribution of the MR from the Hall resistivity is eliminated using the formula $\rho_{yx}(B)= [\rho_{yx}(+B)-\rho_{yx}(-B)]/2$. The obtained non-linear Hall resistivity is shown in Fig.~\ref{Hall}, with a clear signature of both electrons and holes participating in the magnetotransport. The Hall conductivity is obtained from the symmetrized $\rho_{yx}(B)$, using the relation $\sigma_{xy}(B)= \rho_{yx}/(\rho_{xx}^2+\rho_{yx}^2)$. To quantify the contributions of the holes and electrons in the observed Hall conductivity and longitudinal conductivity, we fitted our measured data with the two-carrier model of $\sigma_{xy}(B)$ and $\sigma_{xx}$(B), respectively [see Fig. \ref{Hall}], which is given by \cite{PhysRevB.101.205138}

\begin{equation}
\label{sigma_xy}
\sigma_{xy}(B)=eB\left[\dfrac{n_h\mu^2_{h}}{1+(\mu_{h}B)^2}-\dfrac{n_e\mu^2_{e}}{1+(\mu_{e}B)^2}\right]~,
\end{equation}

\begin{equation}
\label{sigma_xx}
	\sigma_{xx}(B)=e\left[\dfrac{n_h\mu_{h}}{1+(\mu_{h}B)^2}+\dfrac{n_e\mu_{e}}{1+(\mu_{e}B)^2}\right]~.
\end{equation}
Here, $n_h$ and $n_e$ ($\mu_{h}$ and $\mu_{e}$) denote the carrier concentration (mobility) of the holes and the electrons, respectively. The extracted values of carrier concentration and mobility from the fitted curves $\sigma_{xy}$(B) and $\sigma_{xx}$(B) are comparable and exhibit similar temperature dependence as presented in Fig. \ref{Hall_Num}. At low temperatures, the estimated $n_e \sim$ 10$^{17}$ cm$^{-3}$ and $n_h \sim$ 10$^{18}$ cm$^{-3}$, which increase as the temperature rises. On the other hand, at low temperatures  $\mu_{e}$ $\sim$ 10$^2$ cm$^2$V$^{-1}$S$^{-1}$ and $\mu_{h}$ $\sim$ 10$^4$ cm$^2$V$^{-1}$S$^{-1}$. Electron mobility decreases monotonically as temperature increases, whereas hole mobility is nearly temperature independent. The low carrier concentration and relatively large mobility is also observed in several other topological materials \cite{Cd3As2,ZrSiS}. The obtained carrier concentration suggests that SrAl$_2$Si$_2$ is an uncompensated material. Thus, in contrast to other topological systems \cite{ZrSiS,PhysRevB.96.121107,Wang2017}, the large MR in SrAl$_2$Si$_2$ may not arise from electron-hole compensation.

\begin{figure*}
	\includegraphics[width=.9\linewidth]{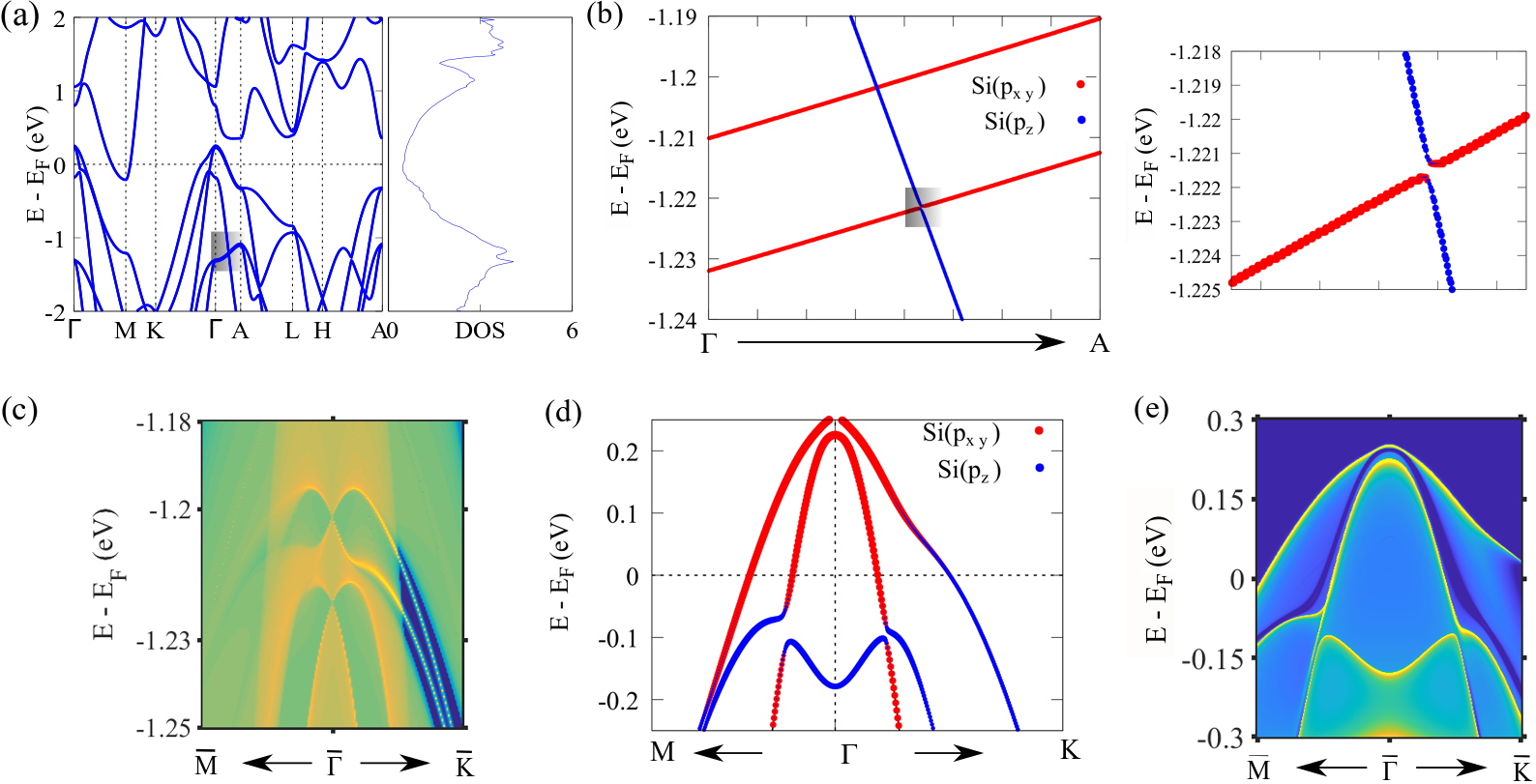} 
	\caption{The band structure and surface states including SOC. (a) The bulk band structure along different high-symmetry directions along
		with the total density of states. (b) Zoomed in band structure along the $\Gamma$-\textit{A} direction with orbital character. The band inversion between Si $p_{xy}$ and $p_z$ orbitals at \textit{E} $\approx$ -1.22 eV is shown separately in the side panel. The Dirac crossing, and the inverted bandgap seen here give rise to topological surface states, as shown in (c). 
		(d) Band inversion near the Fermi energy along \textit{M}-$\Gamma$-\textit{K} direction. Thus the inverted bandgap turns out to be topologically trivial and it is not associated with Dirac like topological surface states, as shown in (e).  
		\label{bandstructur}}
\end{figure*}
\section{ELECTRONIC STRUCTURE}
To understand the electronic properties of SrAl$_2$Si$_2$ better, we now calculate its bulk band-structure in the presence of SOC and it is shown in Fig.~\ref{bandstructur} (a). Clearly, the valence band crosses the Fermi energy near the $\Gamma$ point giving rise to a hole pocket, and the conduction bands crosses the Fermi level in the vicinity of the \textit{M} point creating an electron pocket. A pair of Dirac-like crossings in the bulk bands can be seen along the $\Gamma-K$ direction for $E \approx -0.1$ eV. Another tilted type-I Dirac crossing along the $\Gamma-A$ direction can be observed for $E\approx -1.2$ eV as shown in Fig.~\ref{bandstructur}(b). Moreover, a band inversion between Si $p_{xy}$ and $p_z$ orbitals with a small gap can be observed at \textit{E} $\approx$ -1.22 eV in the zoomed region of Fig.~\ref{bandstructur}(b). 

To explore the topological nature of this band inversion we have  calculated the surface states for the (001) surface. Distinct Dirac like gapless surface states can be observed at \textit{E} $\approx$ -1.20 eV and \textit{E} $\approx$ -1.22 eV, as shown in Fig.~\ref{bandstructur}(c). We also calculate the surface states to explore the band inversion [shown in Fig.~\ref{bandstructur}(d)] near the Fermi energy. However, no distinct surface states arising from the band-inversion in the vicinity of \textit{E} $\approx$ -0.10 eV can be seen in Fig.~\ref{bandstructur}(e), indicating its topologically trivial nature. These features clearly indicates that SrAl$_2$Si$_2$ hosts a type-I topological Dirac semimetal state along with a topological band inversion arising from the band features along the $\Gamma-A$ direction. In addition to these, very clear and high intensity surface states can also be seen in Fig.~\ref{bandstructur}(e), which may arise from a topological band inversion in the conduction band at $E \approx 0.2 $ eV, but which have finite contributions at the Fermi energy. Similar band features have also been observed in other iso-structure compounds \cite{ZrSiS,PhysRevB.96.121107,Wang2017,PhysRevB.101.205138}.

\begin{figure}[H]
	\centering
	\includegraphics[width=0.8\linewidth]{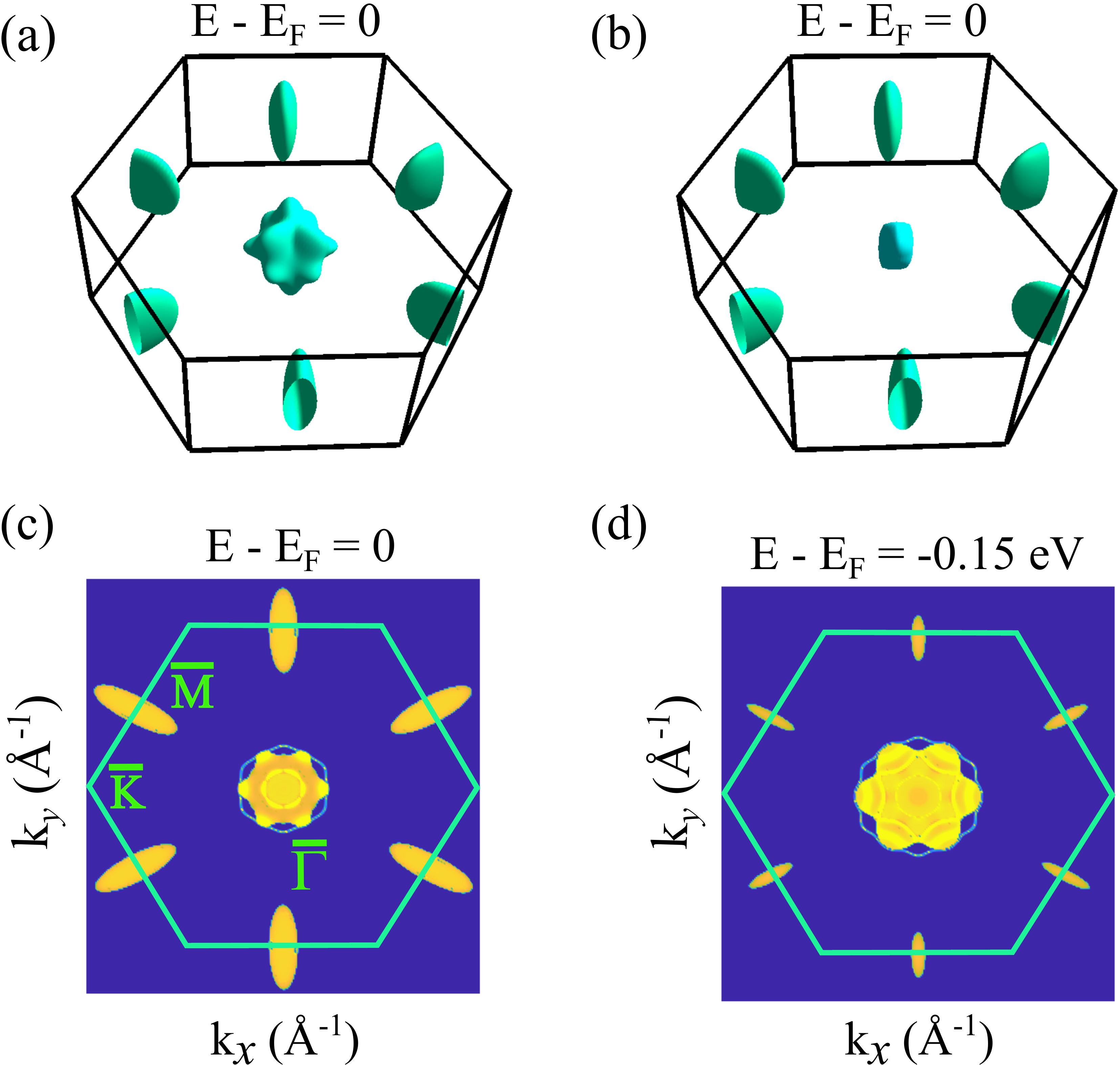} 
	\caption{The bulk and surface projected Fermi surface. (a) The bulk Fermi surface (including SOC) showing the electron pockets around the $M$ point, and a) only the $H_1$ hole pocket and 	b) the smaller $H_2$ hole pocket at the center of the BZ.  (c) -(d) The constant energy contours of the (001) surface projected spectral function for energy values $E - E_F = 0$ and  for $E - E_F = -0.15$ eV, respectively. On decreasing the energy below the Fermi level, the shrinking of the electron pockets, the growth of the hole pockets and the change in the surface	states can be clearly seen.}
	\label{FS}
\end{figure}

Along with these, the band-structure also hosts a pair of hole pocket (labeled $H_1$ and $H_2$) at the $\Gamma$ point, and an electron pocket ($E_1$) at the $M$ point. To explore the structure of these charge carrier pockets, we calculate the Fermi surfaces (with SOC) over the 3D BZ in Fig.~\ref{FS}(a)-(b). A closed set of electron pocket around the $M$ point and a closed pair of hole pocket around the $\Gamma$ point can be clearly seen [pocket $H_1$ in panel a) and pocket $H_2$ in panel b)]. Figs.~\ref{FS}(c)-(d) show the constant energy contours in the (001) surface projected spectral function for different binding energies.  These capture the bulk projected Fermi surfaces, along with the topological surface states. On decreasing the Fermi energy, the hole pockets near the $\Gamma$-point expand whereas the electron pockets at the $M$ point decrease in size, confirming our interpretation. The presence of both the hole and electron pockets is consistent with our two carrier model based fit for the observed Hall resistivity.

\section{DISCUSSION}

Previous measurements reveal that SrAl$_2$Si$_2$ possesses a pseudo-gap with rapidly changing density of states (with energy) in the vicinity of the Fermi energy. Additionally, the sharp band feature hosting the electron and hole pockets can be smeared out by changing the densities of electronic carriers, which have direct influence on the resistivity peak \cite{LUE20111448}. Thus the appearance of a resistivity anomaly can be attributed to the presence of two different charge carriers whose density can change with temperature \cite{KAUZLARICH2009240,LUE20111448}. With the application of the magnetic field, the low-temperature resistivity value gets enhanced, leading to a resistivity plateau. The appearance of a resistivity plateau on the application of a magnetic field is also observed in other topological semimetals \cite{LaSb, PhysRevB.94.041103, ZrSiS, YbCdGe}. The observed large non-saturating MR in SrAl$_2$Si$_2$ is comparable with that observed in other topological semimetals \cite{YSb, CaAgBi, CaCdSn}.

The typical mechanisms that leads to large non-saturation MR in topological semimetals are perfect charge carriers compensation, the nontrivial topological state and a combined effect of a large difference between electron and hole mobility with a moderate carrier compensation  \cite{YSb16}. In our case, the unequal carrier concentration measured from Hall effect rule out the possibility of charge compensation. The theoretically estimated shapes and volumes of the Fermi surfaces also indicate the uncompensated nature of the charge carrier. The uncompensated nature of the charge carriers should reflect in the quadratic variation of MR with $B$. However due to the dominant WAL effect at lower temperature, a quasilinear dispersion is observed. On increasing the temperature, the WAL effect gets suppressed and the dependence of MR moves toward quadratic. There is a type-I Dirac band crossing in SrAl$_2$Si$_2$, but it is far below the Fermi level ($\approx -1.2$ eV). Thus, the contribution of Dirac fermion to the magnetotransport, which is a Fermi surface property, is likely to be insignificant. The large value of MR in SrAl$_2$Si$_2$ is neither related to the charge carrier compensation nor due to non-trivial band topology. Thus, we believe the large MR in SrAl$_2$Si$_2$ 
\noindent arises from combination of moderate charge carrier compensation and substantial difference between electron and hole mobilities. A relatively large MR originating from similar phenomena was reported earlier in other semimetals, such as YSb and SiP$_2$ \cite{YSb16,SiP2}.

\section{CONCLUSIONS}

In conclusion, we have performed a systematic study on a good quality single crystal of SrAl$_2$Si$_2$ by magnetotransport measurement and band structure calculations. Electrical transport data resemble that of metals and semiconductors at low and high temperatures, respectively. The observed large non-saturating MR and resistivity plateaus are like that observed in other topological semimetals. The WAL effect is unveiled by analyzing MC data with the HLN model. The measured  Hall resistivity is nonlinear in $B$, and fits well with the two carrier model, confirming the presence of both electrons and holes as charge carriers. Our band structure calculation suggests that SrAl$_2$Si$_2$ is a topologically non-trivial material with type-I tilted Dirac fermions in addition to other topological band inversions.

\section{ACKNOWLEDGMENTS}

We acknowledge IIT Kanpur, Science and Engineering Research Board (SRG/2019/001686 and CRG/2018/000220), IIT Roorkee (SMILE-13), and the Department of Science and Technology for financial support. We acknowledge the High Performance Computing facility at IIT Kanpur, for computational support.  
	
\bibliography{Reference_SrAl2Si2}
	
\end{document}